\documentclass[prl,aps,twocolumn,preprintnumbers,amsmath,amssymb,nofootinbib,superscriptaddress,notitlepage]{revtex4-1}
\usepackage{epsfig}
\usepackage[utf8]{inputenc}
\usepackage{color}
\usepackage{epstopdf}
\usepackage{multirow}
\usepackage{verbatim}
\usepackage{ftnxtra}
\usepackage{slashed}
\usepackage{ulem}
\usepackage{subfigure}
\usepackage{tikz}

\newcommand{\be}{\begin{eqnarray}}
\newcommand{\ee}{\end{eqnarray}}

\newcommand{\uestc}{\affiliation{School of Physics, University of Electronic Science and
Technology of China, Chengdu 610054, China}}
\newcommand{\ucas}{\affiliation{School of Physical Sciences, University of Chinese Academy of Sciences (UCAS), Beijing 100049, China}}
\newcommand{\kek}{\affiliation{KEK Theory Center, Institute of Particle and Nuclear Studies (IPNS), High Energy Accelerator Research Organization (KEK), 1-1 Oho, Tsukuba, Ibaraki, 305-0801, Japan}}

\newcommand{\asrc}{\affiliation{Advanced Science Research Center, Japan Atomic Energy
Agency, Tokai, Ibaraki, 319-1195, Japan}}

\newcommand{\riken}{\affiliation{Nishina Center
for Accelerator-Based Science, RIKEN, Wako 351-0198, Japan}}

\begin{document}

\title{
A Paradigm for the Coupled-Channel Origin of Resonances: the Exotic $T_{c\bar{s}}$ in $D_{s1}(2460/2536)\to D_s\pi\pi$\\
}

\author{Zhi Yang}\email{zhiyang@uestc.edu.cn}
\uestc

\author{Guang-Juan Wang}\email{wgj@post.kek.jp, corresponding author}
\kek

\author{Jia-Jun Wu}\email{wujiajun@ucas.ac.cn, corresponding author}
\ucas

\author{Makoto Oka}\email{makoto.oka@riken.jp}
\asrc
\riken

\date{\today}

\begin{abstract}

The $T_{c\bar{s}}$ state observed in the decay $D_{s1}(2460)^+ \to D_s^+\pi^+\pi^-$ provides direct evidence for an isovector open-charm tetraquark state with strangeness--a discovery that demands a systematic framework connecting its origin to the nature of the parent $D_{s1}$.
We successfully achieve this connection by two mechanisms, triangle loops and the coupled channel of $DK$–$D_s\pi$ with pure off-diagonal potentials.
We first point out the behavior of propagator of $D_s\pi$ will influence the effective potential of $DK\to DK$, then we can successfully obtain the pole of $T_{c\bar{s}}$ on the second Reimann Sheet.
By combing with the $\pi\pi$-$KK$ rescattering, not only the two-peak structure in $D_{s1}(2460)$ decay is well reproduced, but also a single-peak structure is predicted in $D_{s1}(2536)$ decay.
The marked difference, testable at LHCb and Belle II, is driven by the $S$-wave versus $D$-wave nature of their $D^*K$ couplings, revealing the underlying structural distinction between the two $D_{s1}$ states.
By directly linking hadronic structure to decay patterns, this work provides a template for deciphering the nature of such exotic states. 
More broadly, by revealing how non-perturbative coupled-channel effects manifest in exotic hadrons, our analysis connects to a universal mechanism shared by systems ranging from halo nuclei to atomic Feshbach resonances, offering a unified perspective across these fields.

\end{abstract}

\maketitle

Recent discoveries of the doubly-charged open-charm tetraquark candidate $T_{c\bar{s}}^{++}$ and its neutral partner $T_{c\bar{s}}^{0}$ by the LHCb collaboration~\cite{LHCb:2024iuo} 
provide the first observation of an isovector open-charm tetraquark candidate. 
These states, identified through an amplitude analysis of the $D_{s1}(2460)^+\to D_s^+\pi^+\pi^-$ decay, exhibit a mass of $2327\pm13\pm13$ MeV, a width of $96\pm16^{+170}_{-23}$ MeV, with preferred quantum numbers of $I(J^P)=1(0^+)$~\cite{LHCb:2024iuo}. 
Such a discovery unequivocally transcends the traditional quark model and provides direct evidence for four-quark dynamics. 
It advances the decades-long quest to decipher the non-perturbative dynamics of Quantum Chromodynamics (QCD) through exotic hadron spectroscopy.

A theoretically motivated interpretation is that $T_{c\bar s}$ emerges as a hadronic molecule generated by $DK$-$D_s\pi$ coupled-channel dynamics~\cite{Kolomeitsev:2003ac, Guo:2009ct, Guo:2015dha, Wang:2024fsz}. 
Earlier studies identified a pole near $2300$ MeV on the third Riemann sheet of the $DK$-$D_s\pi$ system~\cite{Guo:2009ct, Guo:2015dha}, while alternative works have suggested that the experimental enhancement can instead arise from a triangle singularity without requiring a genuine resonance~\cite{Roca:2025lij,Dias:2025izv}.
This impasse reflects a deeper methodological gap: rescattering and triangle loops capture complementary facets of the same non-perturbative dynamics, yet no framework has systematically unified them. 
Establishing such a unified framework is not only essential to settle the nature of $T_{c\bar s}$, but also opens the door to connecting the decays of the two $D_{s1}$ states within a single consistent picture---a connection that has remained entirely unexplored.

The $D_{s1}(2460)$, lying just below the $D^*K$ threshold and significantly lighter than the quark-model expectations~\cite{Godfrey:1985xj}, is commonly interpreted as a hadronic molecule~\cite{Kolomeitsev:2003ac,Szczepaniak:2003vy,Hofmann:2003je,vanBeveren:2003kd,Barnes:2003dj,Guo:2006fu,Guo:2006rp,Albaladejo:2018mhb,Wu:2019vsy,Kong:2021ohg,Gregory:2021rgy,Wang:2012bu,Huang:2021fdt} or at least a $c\bar s-D^*K$ mixture with significant molecular component~\cite{Maiani:2004vq,Dai:2006uz}.
By contrast, the $D_{s1}(2536)$ sits close to the quark-model expectations and is naturally regarded as a conventional $c\bar s$ state~\cite{Chen:2016spr,Guo:2017jvc,Chen:2016qju,Esposito:2016noz,Brambilla:2019esw,Kalashnikova:2018vkv}. 
Heavy quark symmetry provides the key to understanding these two states within a unified framework: both originate from two bare $c\bar s$ configurations that couple to $D^*K$ in $S$-wave and $D$-waves, respectively~\cite{Yang:2021tvc}.
The bare state coupled in $S$-wave undergoes a large mass shift, reproducing the molecular-like $D_{s1}(2460)$, while the other predominantly coupled to the $D$-wave $D^*K$ state is only weakly modified, giving rise to the near-conventional $D_{s1}(2536)$. 
This picture successfully reproduces their spectrum and lattice QCD spectra of the $D^{(*)}K$ system~\cite{Lang:2014yfa, Bali:2017pdv}, but its implications for decay observables remain largely unexplored.
Connecting this spectroscopic picture to decay patterns is precisely the missing link needed to transform our understanding of these mysterious‌ states.

The decay kinematics of $D_{s1}^+ \to D_s^+ \pi^+ \pi^-$ act as ``structural fingerprints" that address two fundamental, interconnected questions: 
\begin{itemize}
\item \textbf{Coupled-channel origin of $T_{c\bar{s}}$:} Does the $T_{c\bar{s}}$ resonance arise from the off-diagonal $DK$-$D_s\pi$ dynamics? This universal mechanism is exemplified by atomic Feshbach resonances, controlled by magnetically tuned off-diagonal coupling~\cite{Chin:2010crf}, as well as by the formation of nuclear cluster states, which are governed by the off-diagonal coupling between different cluster channels~\cite{Epelbaum:2012qn,Riisager:2012it}.
\item \textbf{Internal structure of $D_{s1}$:} What can the $D_s \pi \pi$ distributions reveal about the internal structure of $D_{s1}(2460/2536)$? After decades of debate over whether these states are molecules, conventional $c\bar{s}$, or mixtures thereof, their decay kinematics offer a qualitatively new experimental window. The structural contrast between them turns their decay pathways into complementary filters of exotic state formation.
\end{itemize}

\begin{figure}[htb]
\centering
\includegraphics[width=1.0\linewidth]{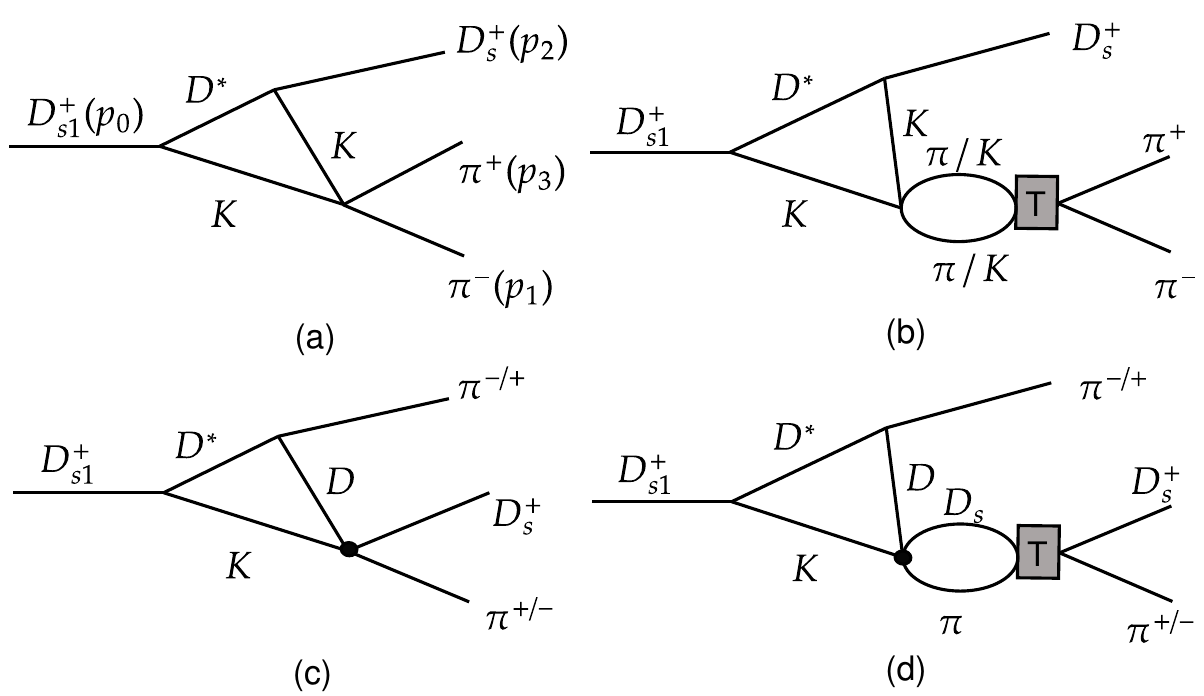}
\caption{The illustrative Feynman diagrams for $D_{s1}(2460)$ and $D_{s1}(2536)$ decays into $D_s^+\pi^+\pi^-$.
}
\label{fig:tcspro}
\end{figure}

In this Letter, we address these questions by introducing a unified framework for the $T_{c\bar{s}}$ state, where both $D_{s1}(2460)$ and $D_{s1}(2536)$ states are treated consistently in their $ D_s\pi\pi$ decays.
For the first time, this enables a direct comparison between the two decay modes and reveals how their structural differences shape observable signals.

To consistently‌ describe the two $D_{s1}$ decays, we construct the $D_{s1}D^*K$ vertex from Ref.~\cite{Yang:2021tvc} and simultaneously incorporate triangle loop and rescattering mechanism, thereby capturing the full dynamics responsible for the $T_{c\bar{s}}$ signal.  
This work provides the first unified description of the decays of both $D_{s1}$ states. 
In particular, the prediction for $D_{s1}(2536)^+ \to D_s^+ \pi^+ \pi^-$ offers an experimental benchmark of the $T_{c\bar{s}}$ dynamics.

The relevant Feynman diagrams are shown in  Fig.~\ref{fig:tcspro}. 
In all cases, the $D_{s1}$ first couples with $D^*K$. 
The virtual $D^*K$ pair then converts into $D_s^+\pi^+\pi^-$ through triangle diagrams,  which
subsequently undergoes $KK-\pi\pi$ and $DK$-$D_s\pi$ coupled-channel rescattering, represented by the box labeled $T$ in Fig.~\ref{fig:tcspro}.
Since both $D_{s1}(2460)$ and $D_{s1}(2536)$ mainly couple to $D^*K$, the triangle diagram provides the primary production mechanism for the three-body final state in both decays.
Importantly, the $D_{s1}D^*K$ couplings used here are fixed from the pole residues of the coupled-channel $T$-matrix analysis in Ref.~\cite{Yang:2021tvc}, which simultaneously describes the spectrum of all four $P$-wave $D_{s}$ excitations together with the lattice-QCD data of the $D^{(*)}K$ system. 
Thus, the present analysis unifies the description of spectroscopy and decay, allowing the $D_s\pi\pi$ decays of both the $D_{s1}$ to be investigated within a single framework, directly linking their internal structure to the dynamics that generate the $T_{c\bar{s}}$.

The effective $D_{s_1}\to D^*K$ vertex in $S$- and $D$-wave can be parameterized as
\begin{equation}
    \mathcal{M}_S=g_S \epsilon_i^\mu \epsilon^{\dagger}_{j, \mu},\quad 
    \mathcal{M}_D=\frac{g_D}{M^2} \epsilon_i^\mu \epsilon_j^{\dagger\nu}H_{\mu\nu}(q),
    \label{eq:gSD}
\end{equation} 
where $H_{\mu\nu}(q)=\left(q_\mu q_\nu-g_{\mu \nu}q^2/4\right)$. 
Here $\epsilon_i$ and $\epsilon^{\dagger}_j$ are the polarization vectors of initial $D_{s1}$ and intermediate $D^*$, respectively. 
$M$ is the $D_{s1}$ mass and $q$ is the relative momentum of the $D^*K$ system. 
The couplings $g_{S,D}$ are determined from the $T$-matrix residues~\cite{Yang:2021tvc}. 

The decay amplitudes for the diagrams in Fig.\ref{fig:tcspro} are 
\begin{widetext}
\begin{eqnarray}
i\mathcal{M}_a &=& r_1 \int \frac{d^{4}q}{(2\pi)^{4}} \frac{\epsilon^{\mu}L_{\mu}}{[q^2-m_K^2][(q+p_2)^2-m_{D^*}^2][(p_0-p_2-q)^2-m_K^2]}, \nonumber\\
i\mathcal{M}_b &=& r_1 \int\frac{d^{4}k}{(2\pi)^{4}} \int \frac{d^{4}q}{(2\pi)^{4}} \frac{\epsilon^{\mu}L_{\mu}G_{\pi\pi/KK}(k,m_{23})T_{\pi\pi/KK\to \pi\pi}(m_{13})}{[q^2-m_K^2][(q+p_2)^2-m_{D^*}^2][(p_0-p_2-q)^2-m_K^2]}, \nonumber\\
i\mathcal{M}_c &=&  r_{2} \int \frac{d^{4}q}{(2\pi)^{4}} \frac{\epsilon^{\mu}N_{\mu}}{[q^2-m_D^2][(q+p_1)^2-m_{D^*}^2][(p_0-p_1-q)^2-m_K^2]}+(p_1
\leftrightarrow p_3), \nonumber\\
i\mathcal{M}_d &=&  r_{2}\int\frac{d^{4}k}{(2\pi)^{4}}\int \frac{d^{4}q}{(2\pi)^{4}} \frac{\epsilon^{\mu}N_{\mu}G_{D_s\pi}(k,m_{23})T_{D_s\pi\to D_s\pi}(\vec{k},\vec{p}_{\text{on}},m_{23})}{[q^2-m_D^2][(q+p_1)^2-m_{D^*}^2][(p_0-p_1-q)^2-m_K^2]}+(p_1
\leftrightarrow p_3),
\label{eq:amp}
\end{eqnarray}
\end{widetext}
where $p_i$($i=1,2,3$) denotes the momenta of the final state mesons (see Fig.~\ref{fig:tcspro}(a)).
The parameters $r_1$ and $r_{2}$ represent the overall couplings in Fig.~\ref{fig:tcspro}(a,b) and Fig.~\ref{fig:tcspro}(c,d), respectively.
$L_{\mu}$ and $N_{\mu}$ are the Lorentz structures of the vertices in the triangle diagrams Fig.~\ref{fig:tcspro}(a,b) and (c,d), respectively.
$\vec{p}_{\text{on}}$ is the three-momentum of $D_s$ in the $D_s\pi$ centre-mass system. 

The amplitude $T_{D_s\pi\to D_s\pi}(\vec{k},\vec{p}_{\text{on}},m_{23})$ is solved by the Lippmann-Schwinger Equation (LSE) as discussed later, where the prescription of the three-momentum reduction is that intermediate $D_s$ and $\pi$ are both on the mass-shell~\cite{Matsuyama:2006rp}.
For the $S$-wave $\pi\pi$ final-state interaction (FSI) in Fig.~\ref{fig:tcspro}(a,b), it can be described by the chiral Lagrangian~\cite{Oller:1997ti,Tang:2023yls} or the phenomenological Lagrangian~\cite{Kamano:2011ih,Wu:2014vma}. Here we adopt the formula in Ref.~\cite{Oller:1997ti} constrained by the measured $\pi\pi/KK\to\pi\pi$ phase shifts~\cite{Cohen:1980cq,Martin:1979gm}.

For the $D_{s1}(2460)\to D^*K$ vertex, 
only the $S$-wave components are retained since the $D$-wave coupling of $D_{s1}(2460)\to D^*K$ is negligible ($g_S/g_D=-0.02$)~\cite{Yang:2021tvc}. 
The Lorentz structures are
\begin{eqnarray}
L_{\mu}&=&P_{\mu\nu}(p_2+q,m_{D^*})(q-p_2)^{\nu} ,\\
N_{\mu}&=&P_{\mu\nu}(p_1+q,m_{D^*})(p_1-q)^{\nu}(p_2-q+2p_3)^{\alpha} \nonumber\\
&&\times P_{\alpha\beta}(p_2-q,m_{K^*})(p_2+q)^{\beta},
\end{eqnarray}
with $P_{\mu\nu}(p,m)=-g_{\mu\nu}+\frac{p_{\mu}p_{\nu}}{m^2}$.

In Fig.~\ref{fig:tcspro}(c,d), the coupled channel effect of the $DK\to D_s\pi$ is illustrated. 
The corresponding $T$-matrix amplitude $T_{DK\to D_s\pi}$ reads,
\begin{eqnarray}
T_{DK\to D_s\pi}=V_{DK\to D_s\pi}(1+G_{D_s\pi}T_{D_s\pi\to D_s\pi}).
\end{eqnarray}
Notably, in the isovector ($I=1$) system, the diagonal interactions vanish ($V_{DK\to DK}=V_{D_s\pi\to D_s\pi}=0$, let us discuss latter), so only the off-diagonal potential $V_{DK\to D_s\pi}$ contributes.
Fig.~\ref{fig:tcspro}(c) corresponds to the tree-level transition, while Fig.~\ref{fig:tcspro}(d) corresponds to the rescattering contribution. 
In the two diagrams, the $V_{DK\to D_s\pi}$ as part of the triangle diagram is encapsulated in the Lorentz structure $N_\mu$, while its interaction strength is absorbed into the coefficient $r_2$.
As we will show later, the amplitude $T_{D_s\pi\to D_s\pi}$, evaluated via unitarized scattering dynamically, generates the $T_{c\bar{s}}$ and reproduces the full $D_s\pi$ spectrum. 
The coupled-channel $T$-matrix can be obtained by solving the relativistic LSE~\cite{Matsuyama:2006rp,Wu:2012md},
\begin{eqnarray} \label{eq:lse}
&&T_{\alpha\beta}(\vec{p},\vec{p}^{\prime};E)={\mathcal
V}_{\alpha\beta}(\vec{p},\vec{p}^{\prime};E)+\sum_{\gamma}\int d\vec{q}
\nonumber\\
&&\,\,\times \frac{{\mathcal
V}_{\alpha\gamma}(\vec{p},\vec{q};E)T_{\gamma\beta}(\vec{q},\vec{p}^{\prime};E)}{E-\sqrt{m_{\gamma_1}^2+q^2}-\sqrt{m_{\gamma_2}^2+q^2}+i\epsilon}.
\end{eqnarray}
Here, the $\alpha, \beta, \gamma=1, 2$ label the coupled channels, with ``1" for $D_s\pi$ and ``2" for $DK$.

The effective potential kernel ${\mathcal V}$ describes the effective interaction in the scattering process $PP\rightarrow PP$ ($P$ is a pseudoscalar meson). 
In one-boson exchange model, the diagonal processes $D_s\pi\rightarrow D_s\pi$ and $DK\rightarrow DK$ proceed via $\rho$ and $\omega$ exchange.
Indeed, the well known Weinberg-Tomozawa interaction can be understood as the zero-range limit of vector-meson exchange~\cite{Guo:2006rp,Liang:2014kra,Wu:2019vsy,Kong:2021ohg}.
However, in the isovector configuration $(I=1)$, these contributions cancel, leaving vanishing diagonal interactions.
This situation differs from the $I=0$ channel, where strong attraction potential of $DK\rightarrow DK$ generates the $D_{s0}(2317)$. 
The new $T_{c\bar{s}}$ is its isovector analogue but with a very different origin.
Consequently, only the off-diagonal transition $D_s\pi\rightarrow DK$ contributes, which is mediated mainly by $K^*$ exchange (the $D^*$ exchange can be absorbed into the coupling constants due to its large mass).
The effective potential is written as
\begin{eqnarray}
{\mathcal V}=\frac{g_{K^*} \left(p_{\pi}+p_K\right)\cdot\left(p_{D_s}+p_D\right)}{(p_{\pi}-p_K)^2 - m_{K^*}^2}\left(\frac{\Lambda_1^{2}}{\Lambda_1^{2}+p_1^{2}}\frac{\Lambda_2^{2}}{\Lambda_2^{2}+p_2^2}\right)^{2}, \nonumber
\end{eqnarray}
where $g_{K^*}$ is the overall coupling constant.
A dipole form factor is introduced to regularize the potential and guarantees convergence.  
The parameters $\Lambda_1$ and $\Lambda_2$ are the cutoffs for the $D_s\pi$ and $DK$ channels, respectively. 
$p_i$ $(i=1,2)$ denote the three-momenta of the hadrons in the rest frame of each channel.

Using the decay amplitude of Eq.~\eqref{eq:amp}, the differential mass distribution is
\begin{equation}
    \frac{d \Gamma}{d m_{13} d m_{23}}=\frac{1}{(2 \pi)^3} \frac{2 m_{13} 2 m_{23}}{32 m_{D_{s 1}}^3} \overline{\sum}\left|\mathcal{M}\right|^2,
\end{equation}
where $\overline{\sum}$ denotes the average over the $D_{s1}$ polarization states. $\mathcal{M}$ is the total amplitude in ~Eq.\eqref{eq:amp}, where the triangle loop integrals are ultraviolet divergent and we use the dimensional regularization within the $\overline{\mathrm{MS}}$ scheme for renormalization.
Analytical manipulations are performed with the FeynCalc package~\cite{Shtabovenko:2023idz}, and numerical evaluations with LoopTools~\cite{Hahn:1998yk}.

The model contains five free parameters, which are determined by fitting to the efficiency-corrected experimental lineshapes. 
Here, we employ pseudo-data generated using the model provided by the LHCb collaboration~\cite{LHCb:2024iuo} to remove detector effects, such as efficiency and background, thereby enabling a direct comparison with the theoretical calculations. 
The number of events is comparable to that reported in Ref.~\cite{LHCb:2024iuo}. 
The solutions are summarized in Tab.~\ref{tab:fitpar}.

\begin{figure*}[t]
\centering
\includegraphics[width=1\linewidth]{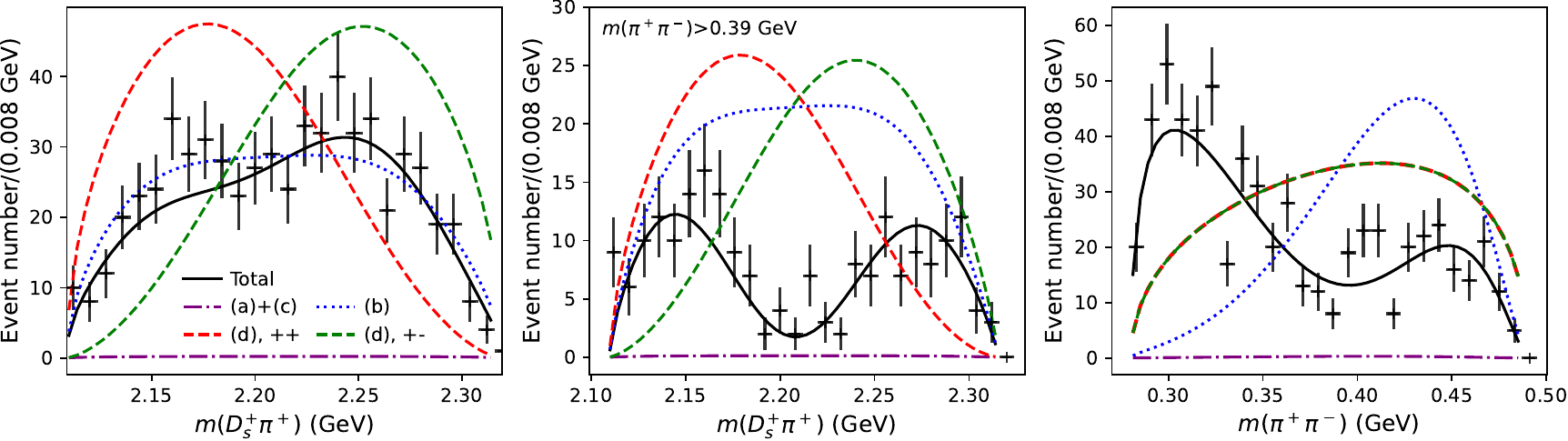}
\caption{The fitted lineshapes of the $T_{c\bar{s}}$ in the $D^+_s\pi^+$ and $\pi^+\pi^-$ invariant mass spectrum. The labels (a), (b), and (c) correspond to the contributions from the individual Feynman diagrams. The "++" and "+-" notations refer to the $D_s^+\pi^+$ and $D_s^+\pi^-$ final states in the $T$-matrix calculation, respectively. Here we use the efficiency-corrected data provided by the LHCb collaboration~\cite{LHCb:2024iuo}.}
\label{fig:fitexp}
\end{figure*}

In Fig.~\ref{fig:fitexp}, we show the contributions of different Feynman diagrams. 
The dominant contributions come from Fig.~\ref{fig:tcspro} (b,d). 
Their interference is crucial to generate the two-peak structure in the $D_s^+\pi^+$ spectrum and governs the $m_{\pi^+\pi^-}$ distribution.
For the diagram (d), the pole in the $D_s^+\pi^+$ channel together with the reflection from the $D_s^+\pi^-$ channel produces two nearby enhancements that tend to overlap into a single broad peak. 
Thus, this diagram alone cannot produce the observed two-peak structure in the $D_s^+\pi^+$ spectrum; the peaks emerge only through the interference with diagram (b). 
Similarly, the $m_{\pi^+\pi^-}$ distribution is shaped by the combined effect of diagrams (b) and (d), leading to a double-bump structure as predicted in Ref.~\cite{Tang:2023yls}, where it was obtained by including the $D^*K$ loop diagrams and considering the $\pi\pi$ FSI.

\begin{figure*}[t]
\centering
\includegraphics[width=0.6\linewidth]{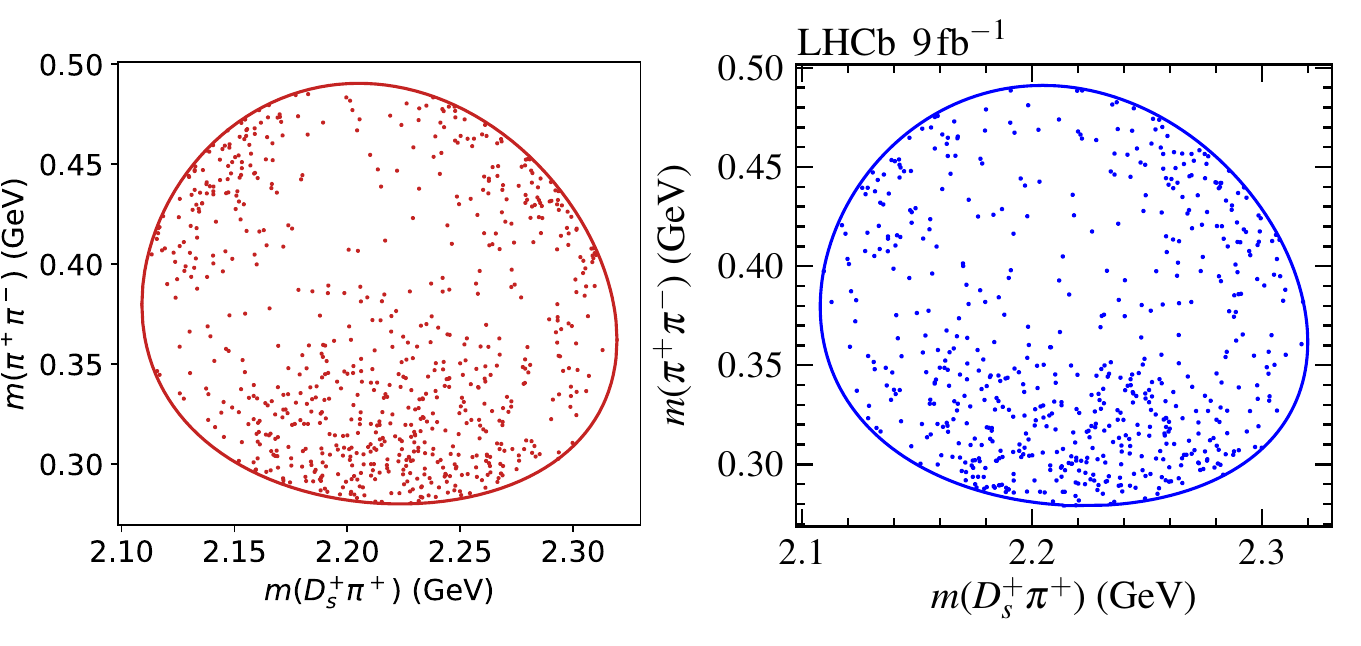}
\caption{The Dalitz plot of the $D_{s1}(2460)^+\to D_s^+\pi^+\pi^-$ process from our fit (left), compared with the detected LHCb result (right).}
\label{fig:dalitz2460}
\end{figure*}

Further support comes from the Dalitz plot shown in Fig.~\ref{fig:dalitz2460}, which is consistent with LHCb data, despite being fitted only to a subset of the invariant-mass spectra.
It reveals a clear depletion in the upper-central area dominated by $\pi\pi$ FSI and enhanced intensity in the lower-central region.
This pattern clearly demonstrates interference beyond the contributions from $\pi\pi$ FSI.

\begin{table}[h]
\centering
\caption{Fitted parameters for Fig. \ref{fig:fitexp}, including $\chi^2/\text{d.o.f.}$ and the pole position $E_p=M-\Gamma/2i$ ( where $M$ and $\Gamma$ denote the mass and width).}
\begin{tabular}{c|cc}
\hline
\textbf{Parameter}  &  \\
\hline
$\Lambda_1$ [GeV]  & $1.60\pm0.07$ \\
$\Lambda_2$ [GeV]  & $0.557\pm0.024$\\
$g_{K^*}$ & $61.4\pm2.4$ \\
$r_1$     & $7.1\pm0.8$ \\
$r_2$     & $0.121\pm0.014$ \\
$\chi^2/\text{d.o.f.}$  & 1.43 \\
\hline
$E_p$ [MeV] & $2211.0^{+31.4}_{-22.6}-154.5^{+20.7}_{-17.1}i$\\
\hline
\end{tabular}
\label{tab:fitpar}
\end{table}

To identify the $T_{c\bar{s}}$ more directly, we solve the Schrödinger equation using the complex scaling method. 
A distinct pole associated with the $DK$–$D_s\pi$ coupled-channel system is found and listed as $E_p$ in Tab.~\ref{tab:fitpar}. 
The pole lies on the second Riemann sheet, characterized by $\operatorname{Im}(q_{D_s\pi}) < 0$ and $\operatorname{Im}(q_{DK}) > 0$, in contrast to the broad third-sheet poles reported in Refs.~\cite{Guo:2009ct, Guo:2015dha}. 
While all analyses indicate the existence of a pole, our result suggests that the second-sheet pole plays the central role in shaping the lineshape.
This discrepancy indicates that the location and nature of the resonance pole are sensitive to the details of the coupled-channel dynamics. Under the heavy-quark symmetry, there exists a pole around $5604.3 - 149.2i$ MeV in the $BK$–$B_s\pi$ coupled-channel system.

In coupled-channel hadron phenomenology, form factor cutoffs are widely regarded as auxiliary parameters irrelevant to dynamical resonance generation, making a uniform cross-channel cutoff standard practice. 
Our work challenges this paradigm.

For $DK \to DK$ (and $D_s\pi \to D_s\pi$) scattering, the effective potential reads $V_{\text{eff}} = \mathcal{V}_{12} G_{D_s\pi} \mathcal{V}_{21}$, with resonant poles determined by
$\det\left(1 - V_{\text{eff}} G_{DK}\right) = 0$.
Pole existence thus depends sensitively on the loop integrals of both channels, not only the off-diagonal potential. 
A shared cutoff leaves the effective potential too weak to produce a dynamical pole; releasing this constraint in our data-driven analysis yields sharply distinct cutoffs, and the $T_{c\bar{s}}$ pole emerges naturally. 
This demonstrates cutoffs are critical to dynamical pole formation in systems with negligible diagonal interactions---a point largely overlooked in prior work.

Channel-dependent form factors have been used phenomenologically \cite{Kamano:2011ih, Ikeno:2021mcb, Wang:2022mxy}, but their microscopic physical origin has remained elusive. We trace this difference to distinct intrinsic momentum scales: the light pion in $D_s\pi$ sustains large virtual momenta in loops, requiring a larger cutoff; the heavier $DK$ system probes smaller momenta, where an oversized cutoff would artificially inflate unphysical high-momentum contributions. This scale disparity directly manifests SU(3) flavor symmetry breaking driven by the strange-light quark mass difference.

\begin{figure*}[t]
\centering
\includegraphics[width=1.0\linewidth]{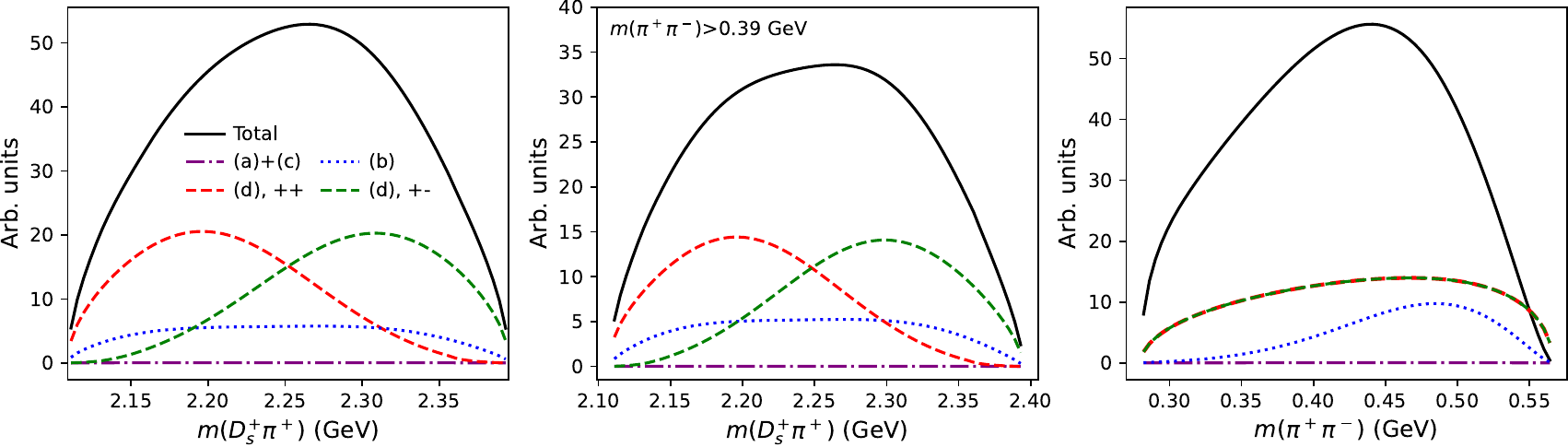}
\caption{The predicted lineshape of the $D_{s1}(2536)^+\to D_s^+\pi^+\pi^-$.}
\label{fig:2536}
\end{figure*}

While our fit establishes the $T_{c\bar{s}}$ resonance pole and highlights the essential role of the off-diagonal couplings, a decisive test requires experimentally verifiable predictions. 
The decay $D_{s1}(2536)^+ \to D_s^+ \pi^+ \pi^-$ serves as such a probe. 
Previous studies of the $D_{s1}(2460)$ and $D_{s1}(2536)$ couplings to $D^*K$ have mainly addressed their mass spectra, leaving their decay properties largely unexplored. 
Within our unified framework, we predict the decay lineshape of $D_{s1}(2536)^+ \to D_s^+ \pi^+ \pi^-$, which offers a direct benchmark for future measurements and a promising opportunity to clarify the structure of the $D_{s1}(2536)^+$.

The decay amplitude of the $D_{s1}(2536)$ is structurally similar to that in Eq.~\eqref{eq:amp}, but requires modification due to its strong coupling to the $D$-wave $D^*K$ channel. 
Experimentally, the $S$- and $D$-wave partial widths for $D_{s1}(2536)\to D^* K$ are comparable~\cite{LHCb:2023eig, ParticleDataGroup:2024cfk}, implying a strong $D$-wave coupling that compensates for the kinematic suppression. 
LHCb extracted the $S$- to $D$-wave amplitude ratio $1.11e^{\pm 0.7i}$ from the decay $B_{(s)}^0 \to D_{s1}(2536)^-K^+ \to \bar{D}^*(2007)^0 K^-K^+$ ~\cite{LHCb:2023eig}, corresponding to $g_S/g_D = 0.1e^{\pm 0.7i}$. 
This is consistent in magnitude with our value $0.08e^{2.7i}$ obtained from the $T$-matrix residues~\cite{Yang:2021tvc}, albeit with a phase difference of approximately $\pi$.
Using either experimental ratios or our theoretical ones leads to very similar invariant-mass distributions. 

\begin{figure}[t]
\centering
\includegraphics[width=0.7\linewidth]{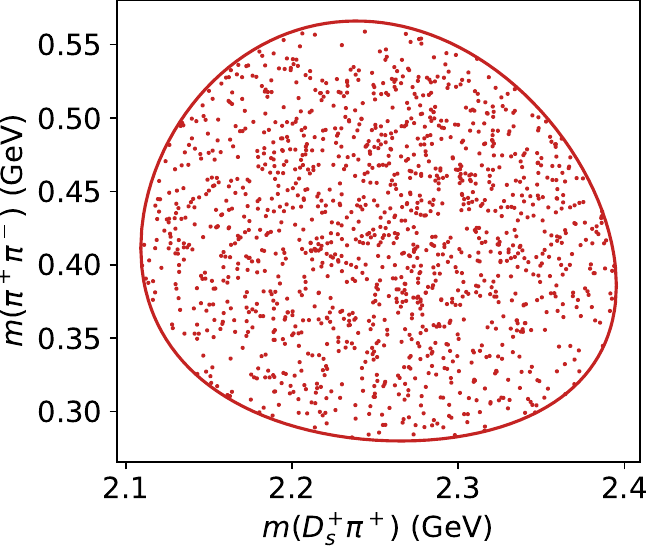}
\caption{The predicted dalitz plot of the $D_{s1}(2536)^+\to D_s^+\pi^+\pi^-$ process.}
\label{fig:dalitz2536}
\end{figure}

Accordingly, both the $S$-wave and $D$-wave couplings must be considered in $D_{s1}(2536)^+ \rightarrow D_s^+ \pi^+ \pi^-$. 
The $S$-wave amplitude follows Eq.~\eqref{eq:amp} with different overall coupling constants $r^{\prime}_{1S}$ and $r^{\prime}_{2S}$ for $L$ and $N$ terms, respectively.
Within the coupled-channel framework, the topologies of the diagrams for $D_{s1}(2460)$ and $D_{s1}(2536)$ are identical, differing only in the $D_{s1}D^*K$ vertices. Thus, we obtain  $r^{\prime}_{1S(2S)}=\lambda r_{1(2)}$.
The $D$-wave coupling of $D_{s1}(2536)$ introduces an additional loop diagram, with the Lorentz structures:
\begin{eqnarray}
L^{\prime}_{\mu}&=&H_{\mu\gamma}(2p_2+2q-p_0)P_{\gamma\nu}(p_2+q,m_{D^*})(q-p_2)^{\nu},\nonumber \\
N^{\prime}_{\mu} &=& H_{\mu\gamma}(2p_1+2q-p_0)P_{\gamma\nu}(p_1+q,m_{D^*})(p_1-q)_{\nu}\nonumber \\
&&\times(p_2-q+2p_3)_{\alpha}P_{\alpha\beta}(p_2-q,m_{K^*})(p_2+q)_{\beta}.\nonumber
\end{eqnarray}
where two overall couplings $r^{\prime}_{1D}$ and $r^{\prime}_{2D}$ are required for $L^\prime$ and $N^\prime$, respectively. 
These are related to the $S$-wave couplings through the ratio $g_D/g_S$ of $D_{s1}(2536)$, $r^{\prime}_{1D(2D)}=r^{\prime}_{1S(2S)}\times g_D/g_S=\lambda r_{1(2)}\times g_D/g_S $. 
Hence, $\lambda$ enters both the $S$- and $D$-wave couplings only as an overall factor, leaving the predicted lineshape of $D_{s1}(2536)^+\to D_s^+\pi^+\pi^-$ unaffected.
The decay can therefore be predicted without introducing additional free parameters, as illustrated in Fig.~\ref{fig:2536} together with the Dalitz plot in Fig.~\ref{fig:dalitz2536}.

We note that compared to $L_\mu^{(\prime)}$, the Lorentz structure of $N_\mu^{(\prime)}$ is more complex, carrying stronger loop-momentum dependence, which diminishes the distinction between the $S$- and $D$-wave couplings at the $D_{s1}D^*K$ vertex.
Then, the contribution of Fig.~\ref{fig:tcspro}(b) for $D_{s1}(2536)$ is quite small compared to that of Fig.~\ref{fig:tcspro}(d).
As a result, the invariant mass spectra in Fig.~\ref{fig:2536} for $D_{s1}(2536)$ decay 
is governed mainly by the pole position of $T_{c\bar{s}}^{++}$ and the reflection from $T_{c\bar{s}}^0$, which yields a broad single enhancement in contrast to the two-peak pattern of $D_{s1}(2460)$ in Fig.~\ref{fig:fitexp}. 
Meanwhile, the $\pi^+\pi^-$ spectrum shows the same features. 
This contrast directly reflects the difference between the two $D_{s1}$ states.
The main observable distinction thus arises from the presence or absence of interference between diagrams. 
This also clarifies why the theoretical and experimental ratios $g_S/g_D$ yield similar line shapes, despite the phase difference.

In this work, we present a unified analysis of $D_{s1}(2460)^+$, $D_{s1}(2536)^+$, and $T_{c\bar{s}}$ by incorporating both the triangle loop and the $DK-D_s\pi$ rescattering based on a unified coupled-channel framework that consistently links spectroscopy and decay. 
%
%
Our main findings are summarized as follows.

First, the $T_{c\bar{s}}$ resonance is dynamically generated by the off-diagonal coupled-channel potential between $DK$ and $D_s\pi$, rather than by diagonal interactions. 
By fitting the LHCb lineshapes, we determine the coupled-channel interactions, which incorporate non-perturbative effects at the
hadronic level and extract a pole corresponding to $T_{c\bar{s}}$ on the second Riemann sheet. Beyond this specific system, our work demonstrates a universal principle: off-diagonal interactions can serve as the primary driver of resonance formation in open quantum systems. This principle directly connects hadron physics to Feshbach resonances in ultracold atoms and cluster formation in nuclear physics, offering a unified conceptual framework that transcends traditional disciplinary boundaries.

Second, the contrasting decay patterns of $D_{s1}$ offer a clean probe of their internal structure: $D_{s1}(2460)$ exhibits the two-peak pattern through the interference between Figs.~\ref{fig:tcspro}(b) and (d), while the $D_{s1}(2536)$ decay is dominated by the rescattering diagram, Fig.~\ref{fig:tcspro}(d). 
Depending on the pole position, our results yield a single broad peak.
We also provide the prediction for the $D_{s1}(2536)^+$ lineshape, which can be directly verified by LHCb and Belle II.

Finally, this study systematically constructs a unified theoretical framework that links the spectroscopy and decay of the $D_{s1}$ states with the dynamics of the $T_{c\bar{s}}$. 
The holistic framework developed here is general and can be extended to analyze other complex hadronic systems, offering a powerful tool for future studies in non-perturbative QCD.

\section{acknowledgments}
\begin{acknowledgments}

We thank Lin-Xuan Zhu, Wen-Bin Qian, Feng-Kun Guo, Wen-Long Sang, Xu-Chang Zheng, Shi-Lin Zhu, Toru Kojo, Akinobu Dote, Bing-Song Zou, Akaki Rusetsky and Satoshi Nakamura for useful discussions and valuable comments. This work is partly supported by the National Natural Science Foundation of China (NSFC) under Grants Nos.~12275046 (Z.Y.), 12175239 and 12221005 (J.J.W), and by the KAKENHI under Grant Nos.~JP20K03959 and JP23K03427 (M.O. and G.J.W), and JP24K17055 (G.J.W), and by the Chinese Academy of Sciences under Grant No. YSBR-101(J.J.W).

\end{acknowledgments}

\bibliography{Tcs.bib}

\end{document}